\documentclass{PoS}

\title{Studies of dark sector \& \boldmath{$B$} decays involving \boldmath{$\tau$} at Belle and Belle II}

\ShortTitle{Studies of dark sector \boldmath{$\&$} B decays involving \boldmath{$\tau$} at Belle and Belle II}

\author{\speaker{Gianluca Inguglia}%
         \thanks{A footnote may follow.}\\
        DESY\\
        E-mail: \email{gianluca.inguglia@desy.de}}

\abstract{The Belle II experiment aims to record 50 ab$^{-1}$ data with the high luminosity to be provided by the SuperKEKB energy-asymmetric $e^+e^−$ collider. The anticipated high statistics data enables us to perform studies of B decays involving $\tau$ leptons such as $B^+\to \tau^+\nu_\tau$ and $B\to D^{(*)}\tau^+\nu_\tau$ modes. The precise measurements of branching fraction and of the $\tau$ lepton polarization in these $B$ decays provide a very sensitive indirect search for a charged Higgs boson. Belle II sensitivity for the charged Higgs is complementary to direct searches at ATLAS and CMS. With the large data sample and by using dedicated triggers the Belle II experiment is expected to explore dark sector by searching for visible and invisible decays of the dark photon and the dark Higgs boson, and by also searching for low mass dark matter with unprecedented precision.}

\FullConference{38th International Conference on High Energy Physics\\
                  3-10 August 2016\\
                 Chicago, USA}

\begin{document}

\section{B decays involving \boldmath{$\tau$}}
\subsection{$B \to \boldmath{\tau \nu}$}
The SM decay rate $\Gamma$ for the decay $B \to \tau \nu$ is calculated in terms of the Fermi constant $G_F$, the mass of the $m_B$ and $m_\tau$ of the $B$ meson and of the $\tau$ lepton, the $B$ meson decay constant and lifetime $f_B$ and $\tau_B$, respectively. This can be written as
\begin{equation}\label{equation0}
\Gamma(B^{+} \to \tau^{+} \nu_\tau)= \frac{G_F^2 m_B m_\tau^2f_B^2}{8\pi}\vert V_{ub}\vert ^2 (1-\frac{m_l^2}{m_B^2})\tau_B.
\end{equation}
New physics beyond the SM can enhance this particular decay. In the supersymmetric two Higgs doublets model (2HDM) the decay rate shown in Eq.~\ref{equation0} assume the form
\begin{equation}
\Gamma(B^{+} \to \tau^{+} \nu_\tau)= \frac{G_F^2 m_B m_\tau^2f_B^2}{8\pi}\vert V_{ub}\vert ^2 (1-\frac{m_\tau^2}{m_B^2})\times (1- \frac{\tan^2\beta}{1+\tilde{ \varepsilon_0} \tan\beta} \frac{m_B^2}{m_{H^\pm}^2})^2
\end{equation}
where $\tan\beta$ defines ratio of the vacuum expectation values (\textit{vev}) for two Higgs doublets and $m_{H^\pm}$ is the mass of the predicted charged Higgs bosons~\cite{chargedhiggs}.
Data collected by the Belle detector in $e^+e^-$ collisions taking place at the mass of the $\Upsilon(4S)$ have been used to search for $B^+ \to \tau^+\nu_\tau$ (and charge conjugated). The $\Upsilon(4S)$ is known to decay to $B^+B^-$ pairs with $BR(\Upsilon(4S)\to B^+B^-)=0.514\pm0.006$, consequently a search for $B^+ \to \tau^+\nu_\tau$ relies on the capabilities of software and detector to (fully) reconstruct one $B$ meson ($B_{rec}$) to infer the flavour of the other ($B_{sig}$). Difficulties in the reconstruction of $B^+\to\tau^+\nu_\tau$ arise mainly from two factors. First the complex algorithms implemented to reconstruct the $B_{rec}$ affects the reconstruction efficiency lowering it, second the $B^+\to\tau^+\nu_\tau$ signal events are characterised by large missing energy due to the presence of neutrinos in the final state (one or two neutrinos in hadronic or leptonic $\tau$ decays, respectively). $B_{rec}$ candidates are selected using the standard variables $M_{bc}=\sqrt{E_{beam}^2-P_B^2}$ and $\Delta E=E_B-E_{beam}$, where $E_B$ and $p_B$ are the energy and momentum of the reconstructed $B$ meson and $E_{beam}$ the beam energy in CM system. Once $B_{rec}$ candidates have been reconstructed, charged tracks are selected to identify the $\tau$ decays for example in the following final states: $\tau^+ \to \mu^+\nu\bar\nu$, $e^+\nu\bar\nu$, $\pi^+\nu$ ,$\pi^+\pi^0\nu$, $\pi^+\pi^+\pi^-\nu$. The signature of the $B\to\tau\nu$ decay is then searched as an excess of events in the $E_{ECL}$ ($=E_{tot}-E_{B_{rec}}-E_{tracks}$) distribution defining the energy deposition in the electromagnetic calorimeter and that is peaked around zero for signal events. A slight excess of events has been observed by the Belle Experiment in two independent analyses using hadronic and semileptonic decays of the $B_{rec}$, and due to limited statistics upper limits have been set to $BR_{B\to\tau\nu}<(1.79_{-0.49}^{+0.56}(stat)_{-0.51}^{+0.46}(syst))$ for hadronic tagged $B_{rec}$~\cite{hadronic} and to $BR_{B\to\tau\nu}<(1.25\pm0.28\pm0.27)\times 10^{-4}$  and $BR_{B\to\tau\nu}<(1.54\pm0.38\pm0.37)\times 10^{-4}$ for semileptonic tagged $B_{rec}$~\cite{semilep} at $90 \%$ confidence level (C.L.). With 50 ab$^{-1}$ of data that will be collected by the Belle II experiment within the next years and with improved detector and software, one would expect to be able to further constrain this decay to $BR_{B^+ \to \tau^+ \nu_\tau}<4\times 10^{-5}$ and to observe at 5 $\sigma$ statistical significance the decay $B^+ \to \mu^+ \nu_\mu$ (in this last case a 5 $\sigma$ observation is expected with just some 5 ab$^{-1}$ of data) as shown in Table.~\ref{my-label}.
\begin{table}[]
\begin{footnotesize}

\begin{center}
%\centering
\caption{Observed (Belle) and expected (Belle II) precision in the determination of  $BR(B^+ \to \tau^+ \nu_\tau)$ for different tagging modes including statistical, systematic, and total uncertainties (in percent).}
\label{my-label}
\begin{tabular}{|c|l|l|l|}
\hline 
\textbf{} Process & Statistical &  Systematic & $Total$ \\ 
					& & (reducible, irreducible) &  \\ \hline \hline 
         $BR(B\to \tau \nu)$ (hadronic tag) &  &  &  \\  
         711 fb$^{-1}$ & 38.0 & (14.2, 4.4) & 40.8 \\  
         5 ab$^{-1}$  & 14.4 & (5.4, 4.4) & 15.8 \\ 
         50 ab$^{-1}$ & 4.6 & (1.6, 4.4) & 6.4 \\ \hline 
         $BR(B\to \tau \nu)$ (semileptonic tag) &  &  &  \\  
         711 fb$^{-1}$ & 24.8 & (18, $_{-9.6} ^{+6.0}$) & $_{-32.2} ^{+31.2}$ \\  
         5 ab$^{-1}$  & 8.6 & (6.2, $_{-9.6} ^{+6.0}$) & $_{-14.4} ^{+12.2}$ \\ 
         50 ab$^{-1}$ & 2.8 & (2.0, $_{-9.6} ^{+6.0}$) & $_{-10.2} ^{+6.8}$ \\ \hline 
         
\end{tabular}
\end{center}
\end{footnotesize}
\end{table}
\subsection{$B \to \boldmath{D^{(*)} \tau \nu}$}
\begin{figure}[!ht]
\begin{center}
\caption{1-sigma contour plot of $R(D)$ vs. $R(D^*)$ showing the results from Babar (black), LHCb (orizzontal cyan band)  Belle (combined semi-leptonic and hadronic tag, blue) and their private preliminary combination (red). In yellow is shown the projection of the the Belle results to Belle II full luminosity (i.e. 50 ab$^{-1}$).}\label{fig:world}
\resizebox{10.5cm}{!}{
\includegraphics{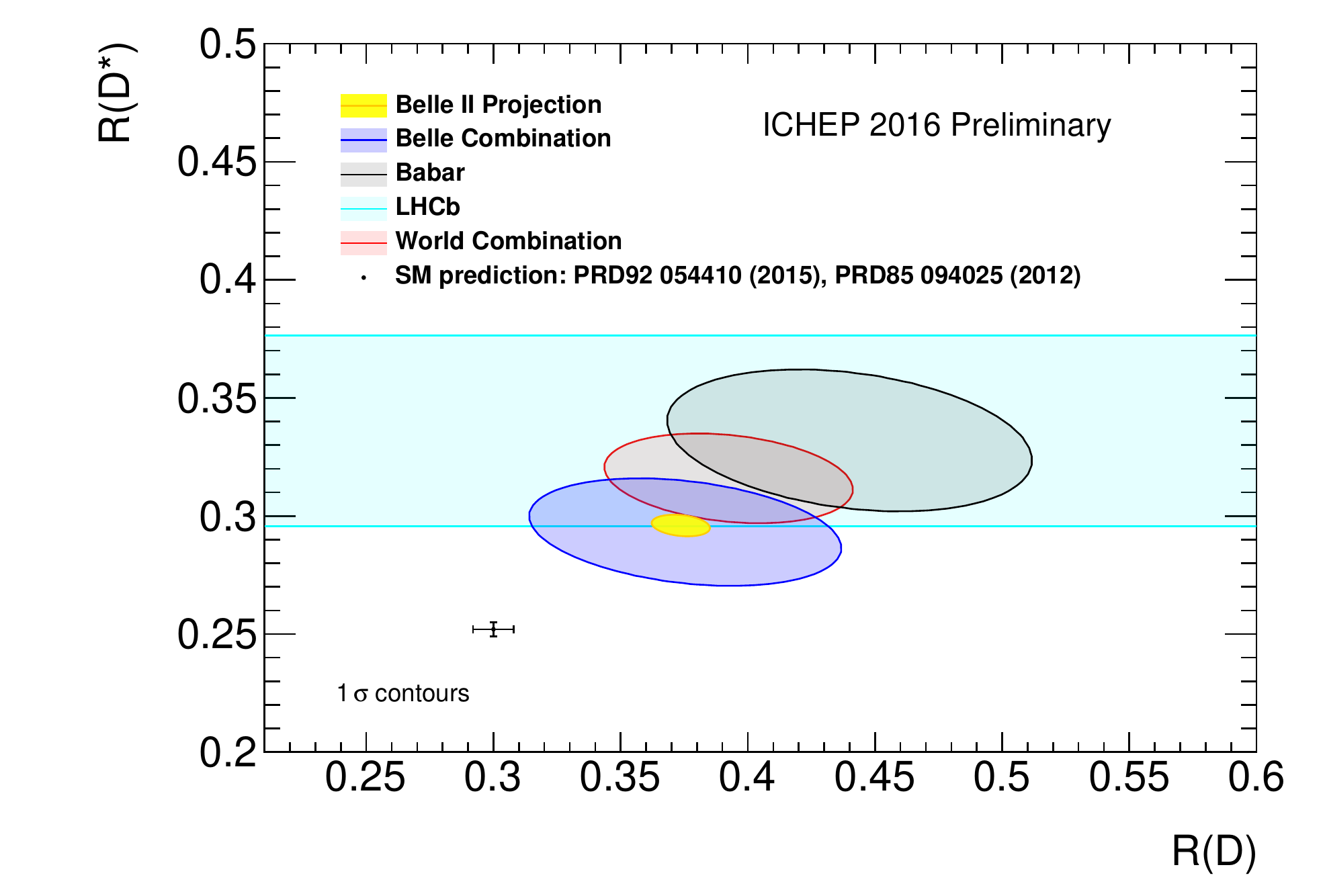}
}
\end{center}
\end{figure}

The semileptonic decay $B \to D^* \tau^+ \nu_{\tau}$ is a $b \to c$ transition proceeding via the emission of a virtual $W^+$ boson in a tree level decay topology. NP can affect this decay in different ways, modifying for example either the BR or the $\tau$ polarization. If NP depends on the mass scale and it is proportional to it as for the case for a charged Higgs boson, in the calculation of branching ratio $B \to D^* \tau^+ \nu_{\tau}$ a term in which the $W^+$ boson is replaced by a charged Higgs boson $H^+$ has to be added, and due to the proportionality to the mass the effect is expected to be more pronounced in decays involving a $\tau$-lepton in the final state with respect to the other two lighter leptons, making such an effect detectable~\cite{2hdm}. An alternative possibility of NP affecting the BR  with respect to the charged Higgs boson is represented by an additional transition (interfering with the SM) in which a virtual leptoquark is produced in the process $b \to \nu_\tau \tilde{h}^*$ and subsequent decay $\tilde{h}^* \to c \tau$~\cite{leptoquark}.\\Two very important quantities that characterise these decay(s) are $R(D)$ and $R(D^*)$ defined as
\begin{equation}\label{eqn:1}
R(D^{(*)})= \frac{\Gamma(B \to D^{(*)} \tau^+ \nu_{\tau})}{\Gamma(B^0 \to D^{(*)} l^+ \nu_{l})_{l=\mu,e}}
\end{equation}
that can be measured experimentally and for which very precise predictions from the SM exist:
\begin{equation}\label{eqn:2}
R(D)=0.297\pm 0.017, 
\end{equation}
\begin{equation}\label{eqn:3}
R(D^*)=0.252\pm 0.003.
\end{equation}
Results reported by the Belle (hadronic tag), $BABAR$ (hadronic tag) and LHCb Collaborations while in agreement with each other have shown a significant deviation from the SM predictions shown in Eqs.~\ref{eqn:2} and \ref{eqn:3} and their averages are $R(D^*)=0.322\pm 0.018\pm 0.012$ and $R(D)=0.391\pm 0.041 \pm 0.028$~\cite{dstar}. The deviation  of the combined results on $R(D)$ and $R(D^*)$ is then found to be 3.9 $\sigma$ from SM prediction. At Belle the strategy for the selection of event candidates proceeds via a two steps process. First an algorithm based on the hierarchical reconstruction of the $B_{rec}$ using NeuroBayes is applied, then a check is performed on the remaining particles seen in the detector to evaluate whether these are consistent with signal signature or not. Recently new results from the Belle Collaboration based on semileptonic decays of the $B_{rec}$ have been released in which $R(D^*)=0.302\pm 0.030 \pm 0.011$ showing an improvement over previously published results~\cite{dstarsl}. The new results are compatible with both the SM and the type-II 2HDM for $\tan\beta/m_H \simeq 0.7$ GeV$^{-1}$ (to be compared to $\tan\beta/m_H \simeq 0.5$ GeV$^{-1}$ obtained for the hadronic tag analysis). Figure~\ref{fig:world} show the mentioned results together with their (private, preliminary) combination and with the projection of Belle results to Belle II luminosities. This decay is also sensitive to the tensor operator in leptoquarks (LQ) models, in particular the $R_2$ LQ model is a good model for compatibility test and assuming $M_{LQ}\simeq 1$ TeV/c$^2$ results show that this model with Wilson coefficient $C_T=+0.36$ is disfavoured. This study is limited by the efficiencies in the full reconstruction and suffer by dominant systematic effects arising from the limited MC sample used for defining the PDF shape and from limited knowledge of probability density function (PDF) shapes in $B\to D^{**}l\nu_l$. One can expect that with larger data and MC samples and with improved software at Belle II one will achieve a large improvement in the determination of $R(D)$ and $R(D^*)$ as shown in Table~\ref{my-label-1}.
\begin{table}[]
\begin{footnotesize}
\begin{center}
\caption{Observed (Belle) and expected (Belle II assuming hadronic tagged $B_{rec}$ mesons.) precision in the determination of $R(D^{(*)})$ including statistical, systematic, and total uncertainties (in percent).}
\label{my-label-1}
\begin{tabular}{|c|l|l|l|}
\hline 
\textbf{} Process & Statistical &  Systematic & $Total$ \\ 
					& & (reducible, irreducible) &  \\ \hline \hline 
         $R(D)$  &  &  &  \\  
         423 fb$^-1$ & 13.1 & (9.1, 3.1) & 16.2 \\  
         5 ab$^-1$  & 3.8 & (2.6, 3.1) & 5.6 \\ 
         50 ab$^-1$ & 1.2 & (0.8, 4.4) & 3.4 \\ \hline 
         $R(D^*)$  &  &  &  \\  
         423 fb$^-1$ & 7.1 & (5.2, 1.9) & 9.0 \\  
         5 ab$^-1$  & 2.1 & (1.5, 1.9) & 3.2 \\ 
         50 ab$^-1$ & 0.7 & (0.5,1.9) & 2.1 \\ \hline 
         \hline
\end{tabular}
\end{center}
\end{footnotesize}
\end{table}

\section{Dark sectors studies}

\subsection{Dark photon}
The dark photon $A'$ is the mediator of a hypothetical dark force related to a $U(1)'$ extension of the SM. Kinetic mixing between the SM photon $\gamma$ and $A'$, $\epsilon F^{Y,\mu\nu}F_{\mu\nu}^D$ with kinetic mixing strength $\varepsilon$ should exist in the interaction Lagrangian and may allow for interactions between SM and dark matter. Dark matter particles would then be neutral under $SU(3)_C \times SU(2)_L \times U(1)_Y$ and charged under $U(1)'_D$ while SM particles would be neutral under $U(1)'$ and the $A'-\gamma$ kinetic mixing term would allow $A'$ to decay to SM particles with very small couplings. Due to the expected low mass of $A'$ (in the range at a few MeV$/$c$^2$ to a few GeV$/$c$^2$~\cite{nima}), $A'$ could be produced in $e^+e^-$ collisions at $B$ and $\tau$-charm factories or in dedicated fixed target experiments in processes that would depend on its mass and its lifetime. In $e^+e^-$ collisions the dark photon is searched for in the reaction $e^+e^- \to \gamma_{ISR} A'$, with subsequent decays of the dark photon to SM final states $A'\to l^+l^-$, $h^+h^-$ ($l=$leptons, $h=$hadrons) or to dark matter $A'\to \chi \bar\chi$, depending on kinematic constraints. The signatures of its production and decays are characterised either by the presence of an energetic photon in the final state plus two oppositely charged tracks with invariant mass equivalent to that of the dark photon or by a mono-energetic photon (for on-shell production). In this case, if $s$ is the CM energy of the collision, the final state is a photon with an energy $E_\gamma= \frac{s-M_{A'}^2}{2\sqrt{s}}$, plus missing energy in the case of decays to dark matter. For SM decays of $A'$ an additional possibility derives from its lifetime. In fact for a short-lived $A'$ one would expect the dark photon to decay promptly near the interaction region but in case of long-lived $A'$ the decay can happen far from the production point. In the latter case the two tracks would form a vertex at a significantly displaced position with respect to the interaction region. Searches for $A'$ are ongoing and are challenged by high background levels (prompt decays), by low trigger efficiencies (displaced decays), and by the need of a single photon trigger (decays into dark matter) that was not available at the Belle experiment, but will be implemented at Belle II. If $A'$ is not observed with the available Belle data, it is possible to anticipate that dark photon decays, to any of the discussed final states, with kinetic mixing $\varepsilon\simeq 10^{-4} $ will be well within reach with the full Belle II data sample.
\subsection{Dark Higgs boson}
A possible way for the dark photon to acquire its mass comes from an extended Higgs sector, In fact the SM Higgs would need to be modified to break the additional $U(1)'$ symmetry. This would lead to a self-interacting dark Higgs boson $h'$ with a mass in the MeV/c$^2$ to GeV/c$^2$ range. In $e^+e^-$ one would search for the dark Higgs boson in a process called dark Higgsstrahlung. Decays of $h'$ depend on its mass: in fact if $M_{h'}>2M_{A'}$ then it would mainly decay to two dark photons, if $M_{A'}<M_{h'}<2M_{A'}$ then $h' \to A' A'^*$ where $A'^*$ decays into leptons, and if $M_{h'}<M_{A'}$ the dark Higgs boson is long lived and it would eventually decay to lepton or hadron pairs. The Belle experiment has searched for the dark Higgs boson in thirteen decay channels, ten of which are exclusive decays such as $2(\pi^+\pi^-)(l^+l^-),3(l^+l^-),2(l^+l^-)(\pi^+\pi^-),3(\pi^+\pi^-)$ and three are inclusive decays such as $2(l^+l^-)X$, where $l=e,\mu$ and $X$ a dark photon decaying invisibly~\cite{TheBelle:2015mwa}. The search for the dark photon and the dark Higgs boson has been performed in the mass ranges of 0.1-3.5 GeV/c$^2$ and 0.2-10.5 GeV/c$^2$, respectively. Since no significant excess of events was observed, individual and combined 90$\%$ C.L. upper limits have been placed on the product of the branching fraction times the Born cross section, $B \times \sigma_{Born}$, and on the product of the dark photon coupling to the dark Higgs boson and the kinetic mixing between the SM photon and the dark photon, $\alpha_D \times \varepsilon^2$. For $\alpha_D=1/137$, $m_{h'} <8$ GeV/c$^2$, and $M_{A'}<$1 GeV/c$^2$, values of the mixing parameter, $\varepsilon$, above $8 \times 10^{−4}$ are excluded at 90$\%$ C.L.~\cite{TheBelle:2015mwa}.
\subsection{Low mass dark matter}
The invisible decay of the $\Upsilon(1S)$ resonance involves neutrino production via $b \bar{b}$ annihilation with BF[$\Upsilon(1S) \to \nu \bar{\nu}] \simeq 10^{-5}$; low mass DM should enhance this BF if $M_{\Upsilon(1S)}>2M_\chi$. The ARGUS, CLEO, Belle and BABAR experiments have searched for invisible decays of the $\Upsilon(1S)$ providing upper limits to BR($\Upsilon(1S) \to invisible$) $<3.0 \times 10^{-4}$ at the 90$\%$ C.L.~\cite{argus,cleo,belle1,babar1}. The Belle II experiment, with a larger data sample than its predecessors thanks to the increased luminosity ($\times 40$ Belle luminosity) and improved detector performance, will be able to further constrain the upper limit on $\Upsilon(1S) \to invisible$ and eventually observe the SM process $\Upsilon(1S)\to \nu \bar{\nu}$ if no new physics is found.\\Due to the disappearance of the $\Upsilon(1S)$ one has to identify tagging algorithms that will be used to unambiguously infer the presence of the $\Upsilon(1S) \to invisible$ decay (these are: di-pion decays of higher spin resonances with data collected at energies equivalent to their mass and via an initial state radiation photon from data collected at the $\Upsilon(4S)$). In addition to the tagging modes for this particular decay channel one has to consider an irreducible peaking background due to two-body decays of the $\Upsilon(1S)$ in which the decay products travel outside of the detector acceptance. One would consider all the tagging methods mentioned before, under the assumption that similar or improved performance of the Belle II detector is expected over previous performance achieved by the Belle experiment. The decays $\Upsilon(nS) \to \pi^+ \pi^- \Upsilon(1S)$ $(n=2,3)$ might be studied with total efficiencies between 10$\%$ and 20$\%$. Table~\ref{yelds} shows the expected yields of $\Upsilon(1S) \to invisible$ for various tagging techniques. It is clear from the yields shown in Tab.~\ref{yelds} the great discovery potential of the Belle II experiment when looking for new physics in the rare $\Upsilon(1S) \to invisible$ decay.
\begin{table}[!ht]
\begin{footnotesize}
\begin{center}
\caption{Expected yields for various $\Upsilon(1S)$ tagging techniques where $L_{int}$ is the integrated luminosity considered for the extrapolation of the yields, $\epsilon$ is the expected total efficiency, $N(\Upsilon(1S))$ is the number of $\Upsilon(1S)$ produced in the process, and $N_{\Upsilon(1S) \to \nu \bar{\nu}}$ and $N_{NP}$ are the expected number of observed $\Upsilon(1S) \to invisible$ events assuming SM ($1\times 10^{-5}$) and new physics ($3\times 10^{-4}$) BF, respectively.}
\label{yelds}
\begin{tabular}{|c|l|l|l|l|l|}
\hline 
\textbf{} Process & $L_{int} (ab^{-1})$ &  $\epsilon$ & $N(\Upsilon(1S))$ & $N_{\Upsilon(1S)\to \nu \bar \nu}$ & $N_{NP}$ \\ \hline \hline 
         $\Upsilon(2S) \to \pi^+ \pi^- \Upsilon(1S)$ & $0.2, \Upsilon(2S)$ & 0.1-0.2 & $2.3\times 10^8$ & 232-464 & 6960-13920 \\ \hline 
         $\Upsilon(3S) \to \pi^+ \pi^- \Upsilon(1S)$ & $0.2, \Upsilon(3S)$ & 0.1-0.2 & $3.2\times 10^7$ & 32-64 & 945-1890 \\ \hline 
         $\Upsilon(4S) \to \pi^+ \pi^- \Upsilon(1S)$ & $50.0, \Upsilon(4S)$& 0.1-0.2 & $5.5\times 10^6$ & 5.5-11 & 165-310 \\ \hline
         $\Upsilon(5S) \to \pi^+ \pi^- \Upsilon(1S)$ & $5.0, \Upsilon(5S)$& 0.1-0.2 & $7.6\times 10^6$ & 7.6-15.2 & 228-456 \\ \hline
         $\gamma_{ISR} \Upsilon(2S) \to (\gamma_{ISR}) \pi^+ \pi^- \Upsilon(1S)$ & $50.0, \Upsilon(4S)$ & 0.1-0.2 & $1.5\times 10^8$ & 150-300 & 4500-9000 \\ \hline
         $\gamma_{ISR} \Upsilon(3S) \to (\gamma_{ISR}) \pi^+ \pi^- \Upsilon(1S)$ & $50.0, \Upsilon(4S)$ & 0.1-0.2 & $6.5\times 10^7$ & 65-130 & 1950-3900 \\ \hline
\hline
\end{tabular}
     \end{center}
\end{footnotesize}

\end{table}

\end{document}